\documentclass[prl,twocolumn,floatfix,showpacs,superscriptaddress]{revtex4-1}
\usepackage{graphics}
\usepackage{epsfig}
\usepackage{times}
\usepackage{bm}
\usepackage{braket}
\usepackage{color}

\usepackage{amsmath}	
\begin{document}
\title{Cross-over from an incommensurate singlet spiral state with 
a vanishingly small spin-gap to a valence bond solid state in
dimerized frustrated ferromagnetic spin-chains}
\author{Cli\`o Efthimia Agrapidis}
\affiliation{Institute for Theoretical Solid State Physics, IFW Dresden, 01069 Dresden, Germany}
\author{Stefan-Ludwig Drechsler}
\affiliation{Institute for Theoretical Solid State Physics, IFW Dresden, 01069 Dresden, Germany}
\author{Jeroen van den Brink}
\affiliation{Institute for Theoretical Solid State Physics, IFW Dresden, 01069 Dresden, Germany}
\affiliation{Department of Physics, Technical University Dresden, 01069 Dresden, Germany}
\author{Satoshi Nishimoto}
 \affiliation{Institute for Theoretical Solid State Physics, IFW Dresden, 01069 Dresden, Germany}
\affiliation{Department of Physics, Technical University Dresden, 01069 Dresden, Germany}

\date{\today}

\begin{abstract}
Motivated by the magnetic properties of the spin-chain compounds 
LiCuSbO$_4$$\equiv$LiSbCuO$_4$ and Rb$_2$Cu$_2$Mo$_3$O$_{12}$,
we study the ground state of the Heisenberg chain with dimerized nearest-neighbor ferromagnetic (FM)
($J_1, J_1^\prime<0$) and next-nearest-neighbor antiferromagnetic ($J_2>0$) couplings.
Using the density-matrix renormalization group technique and spin-wave theory we find a first-order 
transition between a fully-polarized FM
and an incommensurate spiral state at $2\alpha=\beta/(1+\beta)$, where $\alpha$ is the frustration
ratio $J_2/|J_1|$  and $\beta$ the degree of  dimerization $J_1^\prime/J_1$.
In the singlet spiral state the spin-gap is vanishingly small in the vicinity of the FM transition, 
corresponding to a situation of LiCuSbO$_4$.
For larger $\alpha$, corresponding to  Rb$_2$Cu$_2$Mo$_3$O$_{12}$, and smaller $\beta$ there is 
a crossover from this frustration induced incommensurate state to an Affleck-Lieb-Kennedy-Tasaki-type 
valence bond solid state with substantial spin-gaps.
\end{abstract}
\pacs{75.10.Jm, 75.10.Kt, 75.40.Mg}
\maketitle

{\it Introduction.---}
The exotic phenomena emerged by magnetic frustration have long been fascinating
subjects of research in condensed matter physics~\cite{moessner06}. Nowadays,
quasi one-dimensional (1D) frustrated systems, despite their simple structure, are
at the center of attention as a playground for novel ground states that can emerge
from frustration and strong quantum fluctuations due to low dimensionality.
So far, various unconventional magnetic states such as quantum spin liquids~\cite{balents10,helton07},
spin-Peierls states~\cite{arai96}, and Tomonaga-Luttinger (TL) liquid phases~\cite{willenberg15} have been investigated.
Currently, among the hottest topics are magnetic multipolar and in particular spin-nematic
states~\cite{chubukov91,zhitomirsky10,mourigal12,nawa14,buttgen14,nishimoto15}
in which magnon bound states are formed from a subtle competition between geometrical
balance of ferromagnetic (FM) and antiferromagnetic (AFM) correlations among spins.

\begin{figure}[t]
\centering
\includegraphics[width=0.85\linewidth]{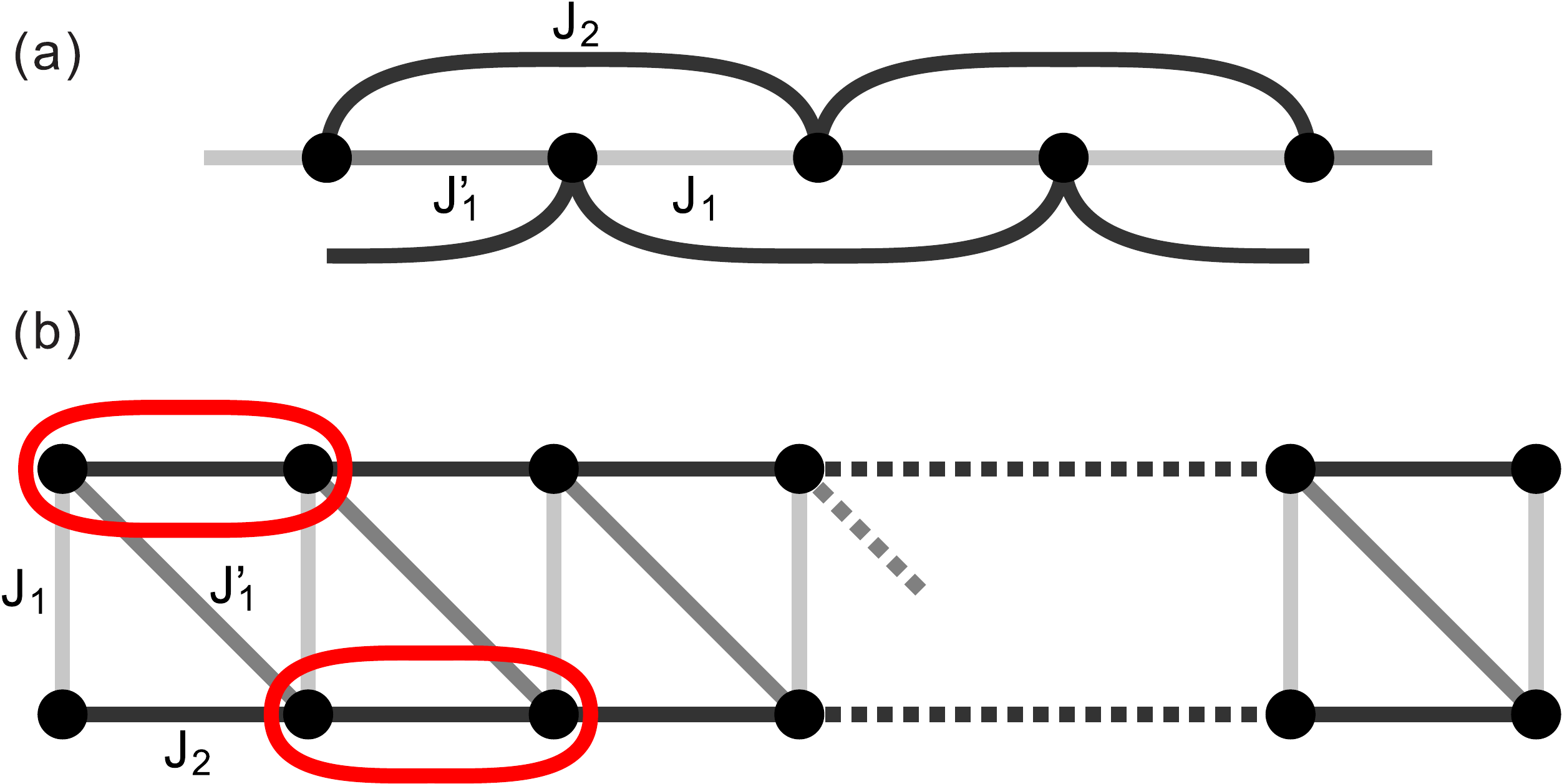}
\caption{
(a) Lattice structure of the $J_1$$-$$J_1^\prime$$-$$J_2$ model. (b) Topologically equivalent
situation which allows to schematic picture of the valence-bond-solid gapped state. Red ellipses
indicate spin-singlet pairs that form in the AKLT (Haldane) state.
}
\label{lattice}
\end{figure}

Very recently, a magnetic field-induced ``hidden'' spin-nematic state was reported in the anisotropic
frustrated spin-chain cuprate LiCuSbO$_4$~\cite{grafe16}. By the nuclear magnetic resonance technique,
a field-induced spin gap was observed above a field $\sim13$T in the measurements of the $^7$Li
spin relaxation rate $T_1^{-1}$, supported by static magnetization and electron spin resonance data.
This material has a unique crystal structure: In the CuO$_2$ chain, four nonequivalent O$^{2-}$ ions
within a CuO$_4$-plaquette give rise to two kinds of nonequivalent ��left�� and ��right�� Cu-Cu bonds
along the chain direction. This gives rise to alternating nearest-neighbor transfer integrals ($t_1 \neq t_1^\prime$).
As a result, a sizable splitting of the two nearest-neighbor FM exchange integrals was estimated:
$J_1 \approx -160$K and $J_1^\prime \approx -90$K, whereas the next-nearest-neighbor AFM coupling
is $J_2 \approx 37.6$K [see Figure~\ref{lattice}(a)]. Another example of a FM dimerized chain compound
is Rb$_2$Cu$_2$Mo$_3$O$_{12}$ which has CuO$_2$ ribbon chains. Here its ribbon chains are twisted,
so that the Cu-Cu distances and the Cu-O-Cu angles are slightly alternating. Accordingly, a small dimerization 
of the nearest-neighbor exchange integrals is expected. Assuming no dimerization, the values of 
the FM nearest- and AFM next-nearest-neighbor exchanges have been estimated as -138K and 51K, 
respectively, by the fitting of susceptibility and magnetization~\cite{hase04}. Besides, a non-magnetic 
ground state with energy gap $E_g\sim1.6$K has been experimentally detected~\cite{yasui14}.
So far, the 1D dimerized AFM Heisenberg has been extensively studied in connection to the celebrated
spin-Peierls compound CuGeO$_3$~\cite{hase93}. In contrast, the dimerized FM case has been hardly
ever discussed. Recently, only the 
weakly dimerized case has been investigated~\cite{ueda14}.
and theoretical studies are definitely required.

Motivated by the above observations, we study a dimerized FM Heisenberg chain with next-nearest-neighbor
AFM couplings by using the spin-wave theory (SWT) and the density-matrix renormalization group (DMRG) method.
The ground-state phase diagram is obtained as a function of dimerization and frustration strengths,
based on the numerical results of total spin, spin gap, spin-spin correlation function, and Tomonaga-Luttinger
(TL) liquid exponent.
We establish the presence of a frustration induced incommensurate singlet state with a spin-gap that is vanishingly small close to
the vicinity to a first order FM transition, corresponding to the situation of LiCuSbO$_4$. Despite the vanishingly small gap,
correlation lengths are comparable to those in the large-gap region in the phase diagram.
For larger $\alpha$ and smaller $\beta$ there is a crossover from this frustration induced incommensurate state to an
Affleck-Lieb-Kennedy-Tasaki-type valence bond solid (realized at $\beta=0$) with substantial spin-gaps.
We also confirm the presence of finite spin gap in the uniform $J_1-J_2$ limit.

{\it Model and method.---} Our spin Hamiltonian is given by
\begin{equation}
H=J_1\sum_{i={\rm even}} {\bf S}_i \cdot {\bf S}_{i+1}+J_1^\prime\sum_{i={\rm odd}} {\bf S}_i \cdot {\bf S}_{i+1}+J_2\sum_i {\bf
S}_i \cdot {\bf S}_{i+2}
\label{ham}
\end{equation}
where ${\bf S}_i$ is a spin-1/2 operator at site $i$. The nearest-neighbor ($J_1, J_1^\prime<0$) and
next-nearest-neighbor ($J_2>0$) interactions are FM and AFM,
respectively [see Figure~\ref{lattice}(a)], and we use the notations of next-nearest-neighbor coupling ratio
$\alpha=J_2/|J_1|$ and nearest neighbor coupling ratio $\beta=J_1^\prime/J_1$ hereafter.

When the system is undimerized ($\beta=1$), we are dealing with the so-called
$J_1-J_2$ model. Increasing $\alpha$, a phase with incommensurate spin-spin correlations 
follows a FM phase. The transition occurs at $\alpha=1/4$, both in the quantum as well as
in the classical model~\cite{bader79,hartel08}. The incommensurate (``spiral'') correlations
are short ranged in the quantum model~\cite{bursill95,nersesyan98}. A vanishingly small gap
was predicted by the field-theory analysis~\cite{itoi01} but no numerical evidence exists so far.
In the limit of $\beta=0$, the system (\ref{ham}) is equivalent to spin ladder with AFM legs and
FM rung couplings. Since this system can be effectively reduced to
an $S=1$ AFM Heisenberg chain with regarding two $S=1/2$ spins on each rung as a $S=1$
spin~\cite{hida91,watanabe93}, the ground state is gapped as predicted by Haldane conjecture~\cite{haldane83}.
Therefore, the ground state can be well described by a valence-bond-solid (VBS)
picture, proposed in the Affleck-Lieb-Kennedy-Tasaki (AKLT) model~\cite{affleck87}.
The schematic picture is shown in Figure~\ref{lattice}(b).

\begin{figure}[t]
\centering
\includegraphics[width=0.95\linewidth]{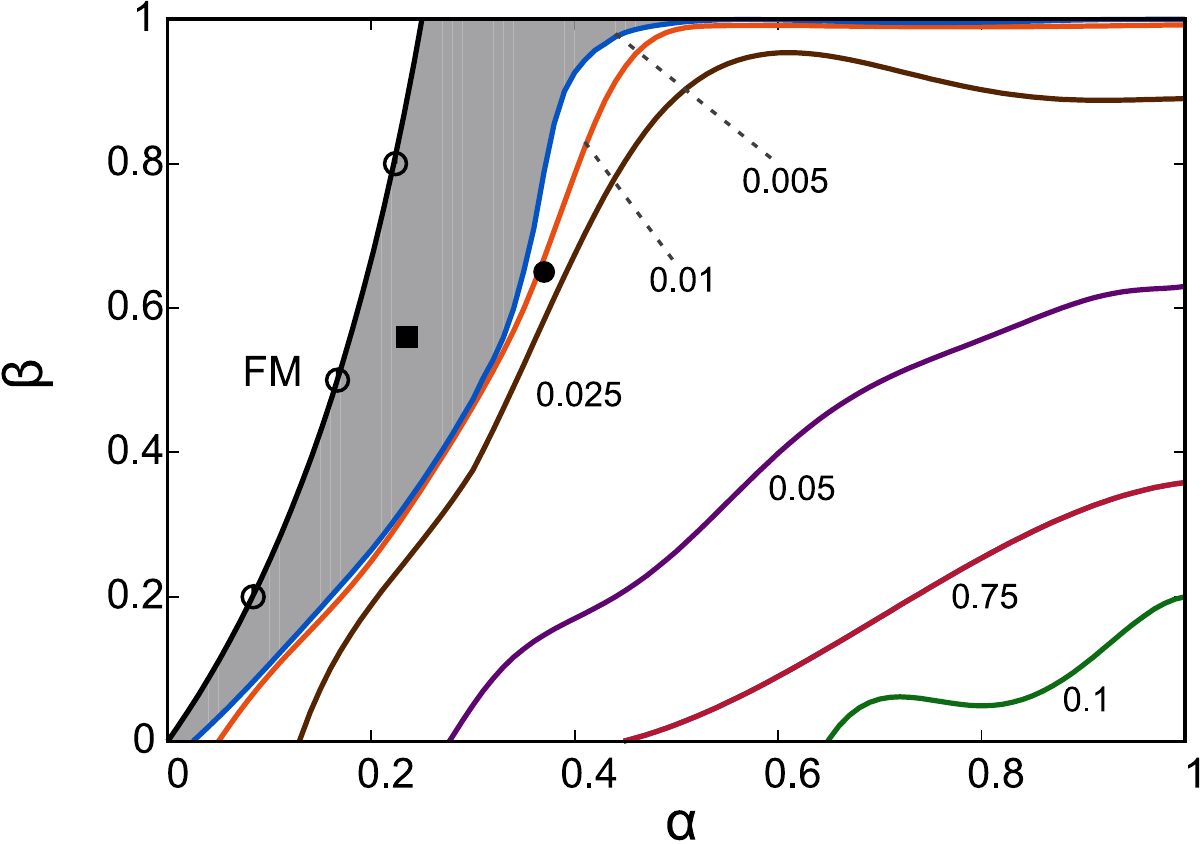}
\caption{
Phase diagram of the $J_1$$-$$J_1^\prime$$-$$J_2$ model in the $\alpha$-$\beta$
plane. Contour map for the spin gap $\Delta/|J_1|$ is shown. The black line represents the boundary of the
fully-polarized ferromagnetic and gapped incommensurate spiral states, obtained by spin-wave
theory. The open circles mark the results from DMRG. The shaded area indicates the region with
a vanishingly small gap ($\Delta|J_1|<10^{-3}$). Filled circle and square indicate the locations of Rb$_2$Cu$_2$Mo$_3$O$_{12}$ and LiCuSbO$_4$, respectively.
}
\label{pd}
\end{figure}

The DMRG method~\cite{white92} is employed to investigate the ground-state properties of
the system (\ref{ham}). We calculate the total spin with periodic boundary conditions, and
spin gap, spin-spin correlation functions, Tomonaga-Luttinger (TL) spin exponent with open
boundary conditions. We keep up to $m=6000$ density-matrix eigenstates in the renormalization
procedure and extrapolate the calculated quantities to the limit $m \to \infty$ if necessary.
Furthermore, several chains with lengths up to $L=800$ are studied to handle the finite-size
effects. In this way, we can obtain quite accurate ground states within the error of
$\Delta E/L=10^{-9}-10^{-10}|J_1|$.

\begin{figure}[!t]
\centering
\includegraphics[width=0.9\linewidth]{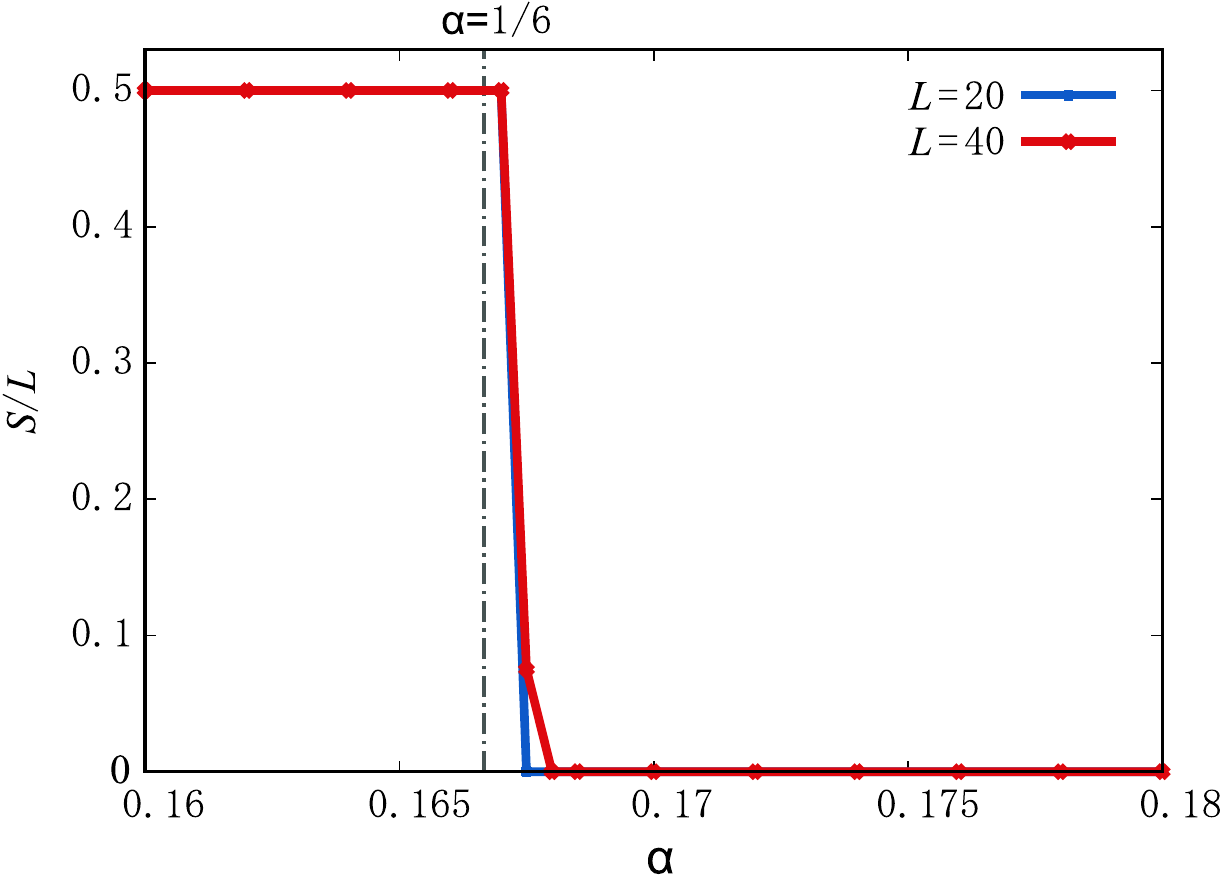}
\caption{Normalized total spin as a function of $\alpha$ at $\beta= 0.5$, calculated by DMRG with periodic boundary conditions.}
\label{spin}
\end{figure}

{\it Ferromagnetic critical point.---}
In the limit of $\beta=0$ and $\alpha=0$, the FM critical point no longer exists because the system is solely composed
of isolated spin-triplet dimers. However, if $\beta$ is finite, the FM order is expected for small $\alpha$.
Let us then consider the $\beta$-dependence of the critical point. Since the quantum fluctuations vanish
at the FM critical point, the classical SWT may work perfectly for estimating the FM critical point.
By the SWT the excitation energy for a FM ground state is given as
$2\omega_q=-\sqrt{1+\beta^2+2\beta\cos(2q)}+2\alpha\cos(2q)$.
The system is in the FM ground state if $\omega_q>0$ for all $q$; otherwise, it is in the spiral
singlet state. Thus, the FM critical point is derived as
\begin{equation}
\alpha_{c,1}=\frac{\beta}{2(1+\beta)}.
\end{equation}
As shown in Figure~\ref{pd}, the FM region is simply shrunk with decreasing $\beta$, and
disappears in the limit of $\beta=0$ as a consequence of isolated FM dimers. It can be numerically confirmed by calculating the ground-state expectation
value of the total-spin quantum number $S$ of the whole system, ${\bm S}^2$, defined as 
$\langle{\bm S}^2\rangle=S(S+1)=\sum_{ij}\langle{\bm S}_i\cdot{\bm S}_j\rangle$. In Figure~\ref{spin}, the normalized total spin at $\beta=0.5$ is plotted as a function of $\alpha$. We can find a direct jump from $S=0$ to $S=L/2$ at $\alpha \sim 0.17$, indicating the absence of an intermediate (partially polarized) FM state. This critical value is in good agreement with that obtained by the SWT ($\alpha_{c,1}=1/6$). Similarly, for all $\beta values$, we confirm direct transition between FM ($S=L/2$) and singlet spiral ($S=0$) states as well as perfect agreement between SWT and DMRG critical points as compared in Figure~\ref{pd}.

{\it Haldane gapped state.---}
So far, the spin gapped state has been verified in the limit of $\beta=0$~\cite{hida91,watanabe93}.
This can be interpreted as a realization of the AKLT VBS state. However, it is a nontrivial question
what happens to the spin gap for finite $\beta$. In our DMRG calculations, the spin gap $\Delta$ is
evaluated as the energy difference between the lowest triplet state and the singlet ground state,
\begin{equation}
\Delta(L)=E_0(L,S^z=1)-E_0(L,S^z=0),\ \ \ \ \ \Delta=\lim_{L\to\infty}\Delta(L),
\end{equation}
where $E_0(L)$ is the ground-state energy for a given number of system length $L$ and
$z$-component of total spin $S^z$.

\begin{figure}[!t]
\centering
\includegraphics[width=0.8\linewidth]{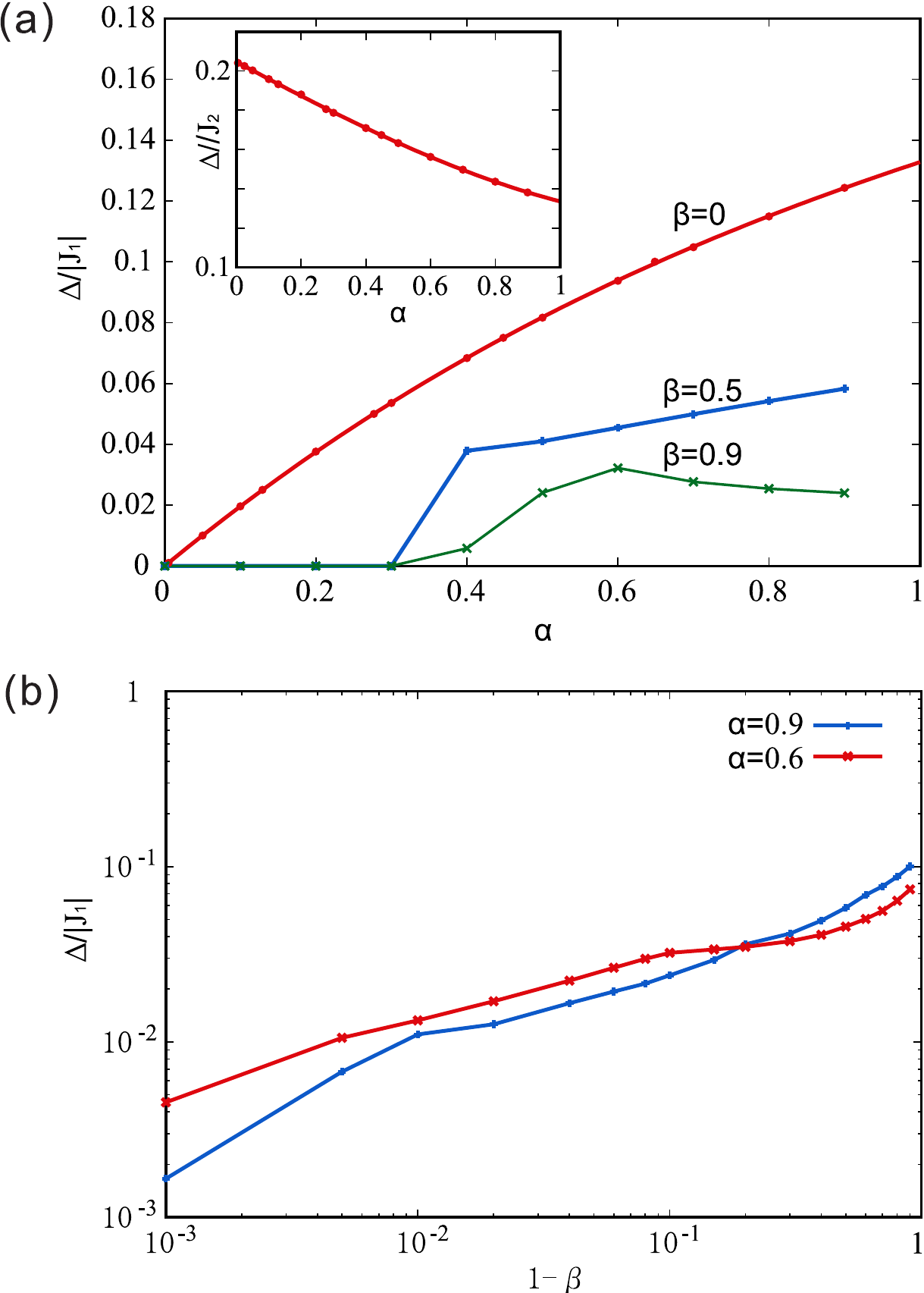}
\caption{
(a) Extrapolated spin gap $\Delta/|J_1|$ as a function of $\alpha$ for $\beta=0$, $0.5$, and $0.9$.
Inset: similar plot of $\Delta/J_2$ at $\beta=0$.
(b) Log-log plot of $\Delta/|J_1|$ as a function of $1-\beta$ for $\alpha=0.6$ and $0.9$.
}
\label{gaps}
\end{figure}

First, we focus on the case of $\beta=0$, namely, a ladder consisting of two AFM leg
chains and FM rungs. In Figure \ref{gaps}(a) the extrapolated values of $\Delta/|J_1|$
is plotted as a function of $\alpha$. The gap opens at $\alpha=0$ and increases monotonously
with increasing $\alpha$, and saturates at a certain value scaled by $|J_1|$. This means that
$\Delta$ is finite for all $\alpha$ at $\beta=0$, which is consistent with the prediction by
the bozonization method~\cite{watanabe93} and the conformal field theory~\cite{timonen91}.
In the limit of $\alpha=0$ the system is exactly reduced to a $S=1$ AFM Heisenberg chain
\begin{equation}
H_{\rm eff}=J_{\rm eff}\sum_i \tilde{\bf S}_i \cdot \tilde{\bf S}_{i+1}-J_1L/4,
\label{effham}
\end{equation}
where $\tilde{\bf S}_i$ is a spin-1 operator as resultant spin
$\tilde{\bf S}_i={\bf S}_{2i}+ {\bf S}_{2i+1}$ and $J_{\rm eff}=J_2/2$.
In the inset of Figure \ref{gaps}(a) $\Delta$ is replotted in unit of $\alpha$.
We obtain $\Delta/J_2=0.2045$ in the limit $\alpha=0$. The Haldane gap of the system
(\ref{effham}) has been calculated as $\Delta/J_{\rm eff}=0.410479$~\cite{ejima15}. Thus,
we can confirm $J_{\rm eff}= J_2/2$ numerically for the mapping from Eq.(\ref{ham}) to
Eq.(\ref{effham}) at the limit $|J_1|/J_2(=1/\alpha)\to 0$ and $\beta=0$.

\begin{figure}[!t]
\centering
\includegraphics[width=0.9\linewidth]{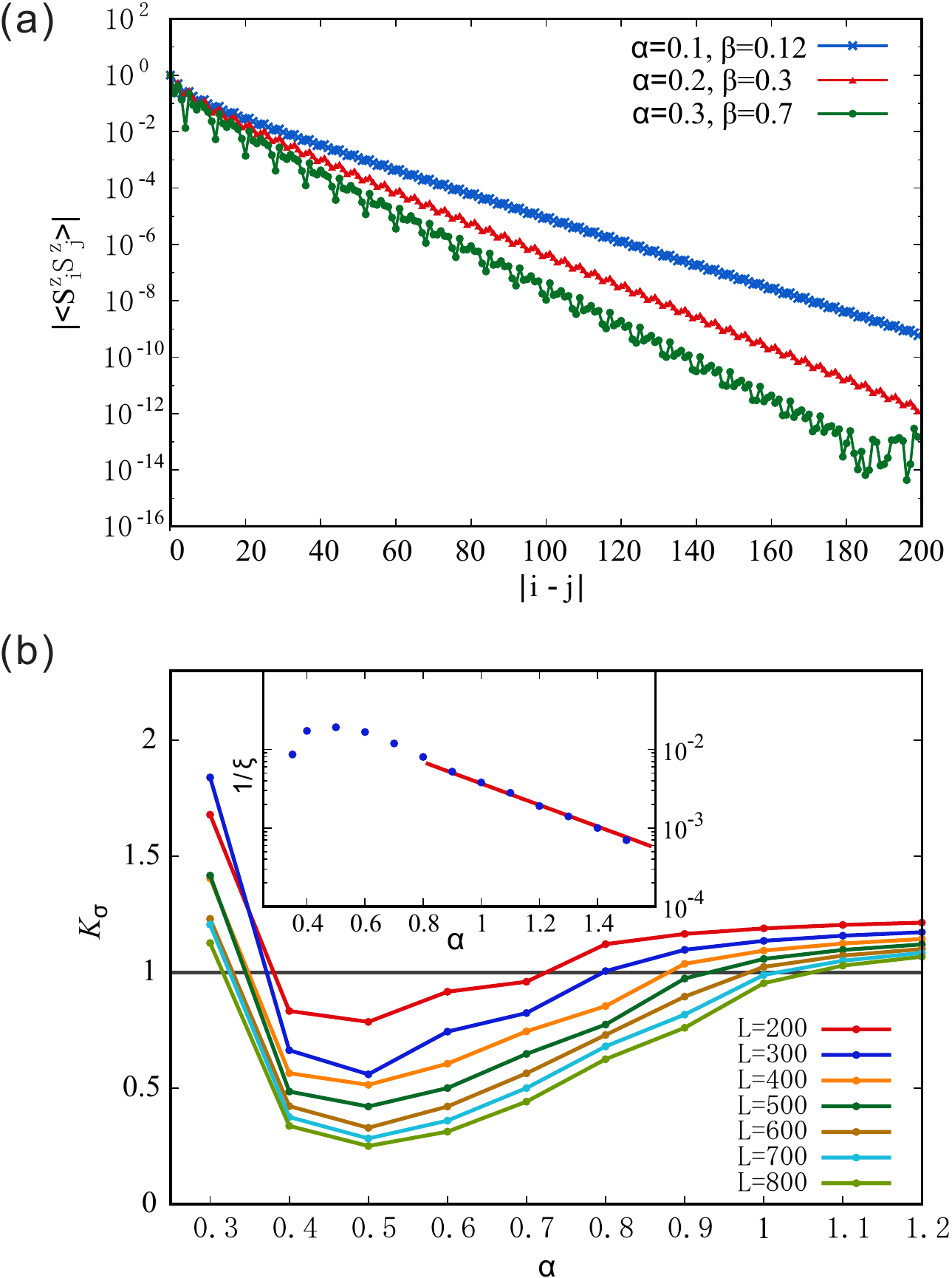}
\caption{
(a) Equal-time spin-spin correlation function $| \langle S^z_i S^z_j \rangle |$ as a function of distance
$|i-j|$ at $\alpha=0.1, \beta=0.12$, $\alpha=0.2, \beta=0.3$, and $\alpha=0.3, \beta=0.7$
for the $L=400$ cluster.
(b) Tomonaga-Luttinger liquid spin exponent as a function of $\alpha$ for systems with several
lengths $L=200-800$. Inset: inverse correlation length $1/\xi$ as a function of $\alpha$. 
The solid line shows a fitting by by $1/\xi=0.085\exp(-\pi\alpha)$.
}
\label{spincorr}
\end{figure}

Next, we look at the effect of $\beta$ on the spin gap. Figure \ref{gaps}(b) shows a log-log
plot of $\Delta/|J_1|$ as a function of $1-\beta$ for $\alpha=0.6$ and $0.9$. The behaviors are
nontrivial but $\Delta$ decays roughly in power law with decreasing $1-\beta$. As a result,
the gap is vanishingly small near the uniform $J_1$$-$$J_2$ limit ($\beta \sim 1$). Besides,
it is interesting that $\Delta$ for $\alpha=0.6$ is larger than that for $\alpha=0.9$
at larger $\beta$ and opposite at smaller $\beta$, which may suggest that
the gapped state near $\beta = 1$ is no longer the AKLT-type VBS state but the frustration
induced one (see below). This is consistent with a maximun gap around $\alpha=0.6$ at
weak dimerization ($\beta=0.9$). On the other hand, an adiabatic connection of
the AKLT-type VBS state from $\beta=0$ to $1$ was predicted by the field-theoretical
analysis for $|J_1|\ll J_2$~\cite{furukawa12}. A contour plot of the magnitude of $\Delta$
is given in Figure~\ref{pd}. We can see a rapid decay of $\Delta$
with approaching the FM phase. However, $\Delta$ is too small to figure out whether it remains
finite, e.g. \ $\Delta \lesssim 10^{-3}$, in the vicinity of the FM critical boundary. Therefore,
to verify the presence or absence of the gap, we checked the asymptotic behavior of
spin-spin correlation function $|\langle S^z_i S^z_j \rangle|$. In Figure \ref{spincorr}(a) the semi-log
plot of $|\langle S^z_i S^z_j \rangle|$ as a function of distance $|i-j|$ is shown for
some parameters near the the FM critical boundary. The distances $|i-j|$ are taken
about the midpoint of the systems to exclude the Friedel oscillations from the system edges,
i.e. $(i+j)/2$ locates around the midpoint of the systems. All of them exhibit exponential
decay of $|\langle S^z_i S^z_j \rangle|$ with distance, which clearly indicates the presence
of a finite spin-gap. The curves are well-fitted with the expression
$| \langle S^z_i S^z_j \rangle | \propto \cos[Q(i-j)]|i-j|^{-\frac{1}{2}}e^{-\frac{|i-j|}{\xi}}$
for long distances~\cite{white96,nomura05}; the correlation lengths $\xi$ are estimated as
$\xi=11.6$ ($\alpha=0.1, \beta=0.12$), $\xi=8.6$ ($\alpha=0.2, \beta=0.3$), and
$\xi=7.3$ ($\alpha=0.3, \beta=0.7$). In the AFM $J_1$$-$$J_2$ model~\cite{white96}, 
a region with $\xi \approx 10$ still have a spin gap of order of $10^{-1}J_1$. 
This may imply the spin velocity of our system is more than two digits smaller than that 
of the AFM $J_1$$-$$J_2$ model since $\Delta=v_s/\xi$ where $v_s$ is the spin velocity.

\begin{figure}[!t]
\centering
\includegraphics[width=0.9\linewidth]{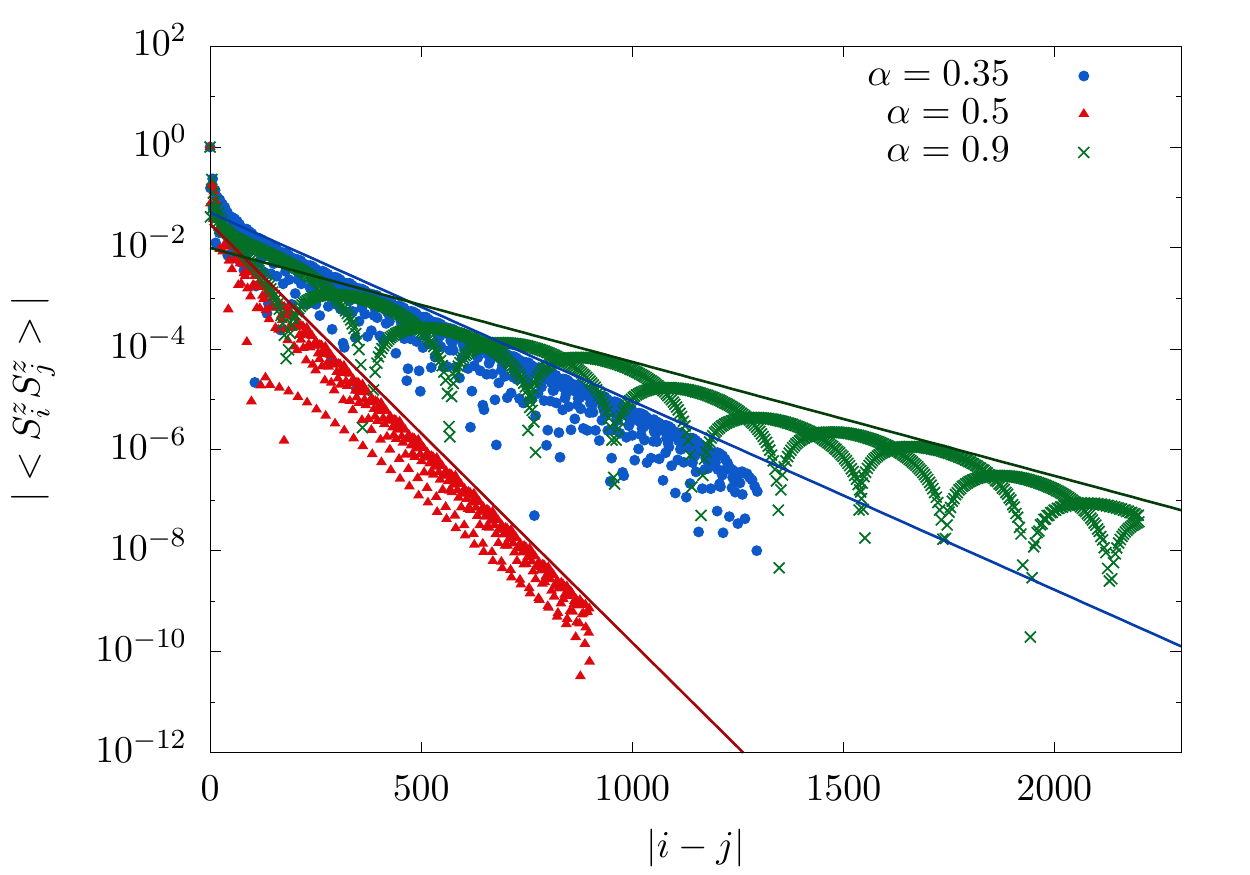}
\caption{Spin-spin correlation functions $|\langle S^z_iS^z_j \rangle|$ as a function of distance $|i-j|$ for several 
$\alpha$ values in the uniform $J_1$$-$$J_2$ model ($\beta=1$). The solid lines exhibit fittings with 
a function $\langle S^z_0 S^z_r \rangle = A \exp(-r/\xi)$ where $\xi$ is the correlation length.
}
\label{expdecay}
\end{figure}

{\it Uniform $J_1$$-$$J_2$ model.---}
In the uniform case ($\beta=1$) the existence of a tiny gap for $\alpha\gtrsim3.3$ was predicted
by the field-theory analysis~\cite{itoi01}. However, the investigation for smaller $\alpha$ is lacking. 
Therefore, to verify the presence or absence of a gap at smaller $\alpha$, we investigated the TL liquid 
spin exponent $K_\sigma$. For our system having four Fermi points ($\pm k_{F1}, \pm k_{F2}$), 
we here assume the asymptotic behavior 
of the spin-spin correlation function to be a power-law decay, like
\begin{equation}
\langle S^z_0 S^z_r \rangle \sim -\frac{K_\sigma}{2\pi^2r^2}+\frac{A\cos[2(k_{F1}-k_{F2})r]}{r^{2K_\sigma}}+\cdots,
\label{spinspin}
\end{equation}
in analogy with the case of two coupled chains~\cite{schulz96}, because the low-energy excitation spectra 
are similar to those of our model~\cite{fabrizio96}. By summing up (\ref{spinspin}) over the distance we obtain
\begin{equation}
K_\sigma=\lim_{L \to \infty}\frac{L}{2}\sum_{kl}e^{i\frac{2\pi}{L}(k-l)}\left\langle S^z_k S^z_l \right\rangle.
\end{equation}
The value of $K_\sigma=0$ indicates a spin-gapped state with an exponential decay of the spin-spin 
correlation in real space; whereas, the convergence to a finite value of $K_\sigma$ in the thermodynamic 
limit suggests a spin-gapless state with the power-law decay ($K_\sigma=1$ within the TL liquid theory). 
In figure~\ref{spincorr}(b), $K_\sigma$ is plotted as a function of $\alpha$ for several chain lengths. 
We clearly find a region where $K_\sigma$ approaches to $0$ with increasing the system size. 
This clearly indicates the existence of a gapped state. The fastest convergence to
$K_\sigma\to0$ around $\alpha=0.5-0.6$ may imply the maximum gap there, similarly to
the case of AFM $J_1$$-$$J_2$ chain. For $\alpha>1$, $K_\sigma$ may seem to converge to 
$K_\sigma=1$. Nevertheless, the validity of the TL liquid theory  is not straightforward around 
$J_2/|J_1|\sim1$, and it is also difficult to exclude the logarithmic corrections for small-gap region. 
Therefore, to consider the connection to the gapped state with tiny gap $\Delta \lesssim 10^{-40}J_2$ 
at $\alpha>3.3$ predicted by the field theory~\cite{itoi01}, we estimated the correlation length 
fas shown in figure~\ref{expdecay}, where the absolute values of spin-spin correlation functions 
$|\langle S^z_iS^z_j \rangle|$ for $\alpha=0.35$, $0.5$, and $0.9$ are plotted as a function of distance $|i-j|$. 
We can clearly see the exponential decays for all $\alpha$ values. By performing the fitting of 
$|\langle S^z_iS^z_j \rangle|$ with a function $\langle S^z_0 S^z_r \rangle = A \exp(-r/\xi)$, where $\xi$ is the correlation length, we obtained 
$\xi=116$ for $\alpha=0.35$, $\xi=52$ for $\alpha=0.5$, and $\xi=192$ for $\alpha=0.9$. In the inset of Figure~\ref{spincorr}(b) the inverse correlation length 
is plotted as a function of $\alpha$. We found that the inverse correlation length is well fitted 
by $1/\xi=0.085\exp(-\pi\alpha)$ for large $\alpha$. Since $\Delta=v_s/\xi$, it may be feasible 
to speculate that the gap has a maximum around $\alpha=0.5-0.6$, decreases with increasing 
$\alpha$, and smoothly connects to the tiny gap region.

Finally, let us explicitly address the relevance of the calculations above for the two spin-chain 
materials mentioned in the introduction. For LiCuSbO$_4$, $\alpha=0.235$ and $\beta=0.56$
are estimated from the density-functional calculations: $J_1 \approx -160$K, $J_1^\prime \approx -90$K,
and $J_2 \approx 37.6$ K~\cite{grafe16,remark}. The system is in the gapped spiral state, 
but very close to the FM phase where the spin-gap is vanishingly small.
Thus, the spin gap may be too small to be detected experimentally. The second compound is
Rb$_2$Cu$_2$Mo$_3$O$_{12}$. If we use the previously estimated parameters $J_1= -138$K and $J_2=51$K
($\alpha=0.37$), a substantial dimerization ($\beta=0.65$) of $J_1$ and $J_1^\prime$ is necessary to
reproduce the experimentally observed gap $E_g\sim1.6$K, namely, $J_1=-138$K and $J_1^\prime=90$K.
Furthermore, if it is more appropriate to consider  the value $-138$K as an averaged FM coupling  $(J_1+J_1^\prime)/2$, then
an even larger dimerization would be needed. In practice, the actual $J_1$ should be somewhat smaller or
$J_2$ should be larger. A detailed analysis of the experimental data that explicitly takes into account 
the dimerization can clarify this point. In the context of these two compounds and also
in general the influence of an external magnetic field is of considerable interest and will be addressed elsewhere. 

{\it Conclusion.---}
%
We considered a frustrated $J_1$$-$$J_2$ spin chain with/without
dimerization of nearest-neighbor FM coupling and determined its phase diagram.
The FM critical point was analytically determined to be $\alpha_c=(\beta/2)/(1+\beta)$ 
by applying the linear spin-wave theory,
which was confirmed by the numerical calculation of the total spin.
The transition between the fully polarized FM and the singlet spiral states is of the first order 
and no partially polarized FM state exists. The spin-gap in the vicinity of the FM boundary 
was confirmed to be finite by the exponential decay of the spin-spin correlation functions 
but it is vanishingly small. In the uniform $J_1$$-$$J_2$ chain, 
the gapped state appears at least around $\alpha \sim 0.5-0.6$
where the TL liquid exponent $K_\sigma$ goes to $0$ in the thermodynamic limit.
Near $\beta=0$ the spin-gap increases with increasing $\alpha$; whereas, near $\beta=1$ it has
a maximum value around the strongest frustration region $\alpha=0.5-0.6$. Therefore, the gap opening
in the entire incommensurate singlet phase may be interpreted as a crossover from the AKLT-type
valence bond solid state near $\beta=0$ to the frustration-induced dimerized state near $\beta=1$.

{\it Acknowledgements.---}
We thank H.\ Rosner, R.\ Kuzian, and J.\ Richter for useful discussions.
We thank U. Nitzsche for technical assistance.
This work is supported by the SFB 1143 of the Deutsche Forschungsgemeinschaft.


\begin{thebibliography}{99}
\bibitem{remark} Thereby
fixed FM contributions to the total $J_1$ and $J'_1$-values have been assumed,
i.e.\  taking into account 
the modification of the AFM superexchange contributions to the former, only. 
An LDA+$U$ analysis available in future
might provide somewhat refined numbers. Within a $pd$-multiband Hubbard 
model this corresponds to fixed direct ferromagnetic exchange interactions
$K_{pd}$ and ferromagnetic Goodenough-Kanamori exchange interactions
involving the Hund's rule coupling involving the intermediate
oxygen sites generic for edge-shared CuO$_2$ chain compounds. The latter might be checked by advanced quantum 
chemistry calculations for small chain clusters. In the absence of results
such calculations at present, the model considered here provides a convenient
tool to investigate phenomenologically LiSbCuO$_4$ and other related compounds.
\end{thebibliography}
\end{document}